\documentclass[english]{article}
\usepackage[T1]{fontenc}
\usepackage{array}
\usepackage{verbatim}
\usepackage{rotating}
\usepackage{url}
\usepackage{multirow}
\usepackage{amstext}
\usepackage{amssymb}
\usepackage{graphicx}

\makeatletter

\providecommand{\tabularnewline}{\\}

\usepackage{ijcai17}

\usepackage{times}

\@ifundefined{showcaptionsetup}{}{%
 \PassOptionsToPackage{caption=false}{subfig}}
\usepackage{subfig}
\makeatother

\usepackage{babel}
\title{End-to-End Prediction of Buffer Overruns from Raw Source Code via
Neural Memory Networks}
\author{Min-je Choi {\normalfont and} Sehun Jeong {\normalfont and} Hakjoo Oh {\normalfont and} Jaegul Choo \\ 
Department of Computer Science and Engineering \\
Korea University, Seoul  \\
\{devnote5676, gifaranga, hakjoo\_oh, jchoo\}@korea.ac.kr}
\begin{document}

\maketitle
\begin{abstract}
Detecting buffer overruns from a source code is one of the most common
and yet challenging tasks in program analysis. Current approaches
based on rigid rules and handcrafted features are limited in terms
of flexible applicability and robustness due to diverse bug patterns
and characteristics existing in sophisticated real-world software
programs. In this paper, we propose a novel, data-driven approach
that is completely end-to-end without requiring any hand-crafted features,
thus free from any program language-specific structural limitations.
In particular, our approach leverages a recently proposed neural network
model called memory networks that have shown the state-of-the-art
performances mainly in question-answering tasks. Our experimental
results using source code samples demonstrate that our proposed model
is capable of accurately detecting different types of buffer overruns.
We also present in-depth analyses on how a memory network can learn
to understand the semantics in programming languages solely from raw
source codes, such as tracing variables of interest, identifying numerical
values, and performing their quantitative comparisons.
\end{abstract}

\section{Introduction}

\label{sec:intro} 

Detecting potential bugs in software programs has long been a challenge
ever since computers were first introduced. To tackle this problem,
researchers in the domain of programming languages developed various
techniques called static analysis, which tries to find potential bugs
in source codes without having to execute them based on a solid mathematical
framework~\cite{cousot1977abstract}. However, designing a static
analyzer is tightly coupled with a particular programming language,
and it is mainly based on a rigid set of rules designed by a few experts,
considering numerous types of possible program states and bug cases.
Thus, even with its slight syntax changes frequently found in real-world
settings, e.g., several variants of ANSI C languages, a significant
amount of engineering effort is required to make a previously designed
analyzer applicable to the other similar languages. 

To overcome these limitations, one can suggest data-driven, machine
learning-based approaches as the rapid growth of deep neural networks
in natural language processing has proved its effectiveness in solving
similar problems such as defect predictions. Studies show that deep
convolutional neural networks (CNNs) and recurrent neural networks
(RNNs) are capable of learning patterns or structures within text
corpora such as source codes, so they can be applied to programming
language tasks such as bug localization~\cite{lam16buggy}, syntax
error correction~\cite{bhatia16auto,pu16skp}, and code suggestion~\cite{white15toward}.

Despite their impressive performances at detecting syntax-level bugs
and code patterns, deep neural networks have shown less success at
understanding how data values are transferred and used within source
codes. This semantic level of understanding requires not only knowledge
on the overall structure but also the capability to track the data
values stored in different variables and methods. Although the aforementioned
deep learning models may learn patterns and structures, they cannot
keep track of how values are changed. This restriction greatly limits
their usefulness in program analysis since run-time bugs and errors
are usually much more difficult to detect and thus are often treated
with greater importance.

In response, we introduce a new deep learning model with the potential
of overcoming such difficulties: memory networks~\cite{weston15mem,sukhbaatar15mem}.
Memory networks are best described as neural networks with external
memory `slots' to store previously introduced information for future
uses. Given a question, it accesses relevant memory slots via an attention
mechanism and combines the values of the accessed slots to reach an
answer. While long short-term memories (LSTMs) and earlier models
also have external memories, theirs tend to evolve as longer sequences
of information are fed in to the network, thus failing to fully preserve
and represent information introduced at earlier stages. Memory networks
on the other hand can preserve the given information even during long
sequences.

This unique aspect of memory networks makes it and its variant models~\cite{kumar16dmn,henaff16ent}
perform exceptionally well at question answering tasks, e.g., the
Facebook bAbI task~\cite{weston15babi}, a widely-used QA benchmark
set. The structure of these tasks comprises a story, a query, and
an answer, from which a model has to predict the correct answer to
the task mentioned in the query by accessing relevant parts of the
given story. These tasks are logical questions such as locating an
object, counting numbers, or basic induction/deduction. All questions
can be correctly answered by referring to appropriate lines of the
given story.

We point out that this task setting is in fact similar to that of
a buffer overrun analysis that requires the understanding of previous
lines in a source code to evaluate whether a buffer access is valid.
Both tasks require knowledge not only on how each line works but also
on how to select the best relevant information from previous lines.
It is this very situation at which our work sets a starting point. 

In this study we set the objective as demonstrating a data-driven
model free of hand-crafted features and rules, and yet capable of
solving tasks with the complexity of buffer overrun analyses. We present
how memory networks can be effectively applied to tasks that require
the understanding of not only syntactic aspects of a source code but
also more complex tasks such as how values are transferred along code
lines. We present how our models can learn the concept of numbers
and numerical comparison simply by training on such buffer overrun
tasks without any additional information. We also introduce a generated
source code dataset that was used to compensate for difficulties we
faced in our data-driven approach. As far as our knowledge goes, our
proposed approach is the first to use deep learning to directly tackle
a run-time error prediction task such as buffer overruns. 

In Section 2, we cover previous work related to our task. In Section
3, we redefine our tasks, introduce our generated dataset and its
purposes, and propose characteristics of the memory network model
and how it is applied to this domain. In Section 4, we report experimental
results and further discuss the performance of memory networks and
notable characteristics it learned during the process. In Section
5 we conclude our work and discuss future work as well as the potential
of memory networks for future tasks.

\section{Related Work}

\label{sec:related work} 

To improve traditional static analysis techniques in the programming
language domain, data-driven approaches based on machine learning
have been recently studied. Obtaining general properties of a target
program, namely, invariants, is one of the prime examples. When concrete
data of target programs such as test cases or logs are available,
data-driven approaches can be used to identify general properties~\cite{sharma2012interpolants,sharma2013verification,sharma2013data,sankaranarayanan2008mining,sankaranarayanan2008dynamic,nori2013termination},
similar to static analysis techniques. This use case is particularly
useful when a target program has inherent complexity that makes contemporary
static analyzers to compromise either of precision and cost, but is
bundled with test cases that can cover most of cases.

Meanwhile, following the upsurge in the developing field of neural
computing and deep learning, many models have been applied to natural
language texts, especially in identifying language structure and patterns.
Socher et al. \cite{socher13rnn} introduced recursive neural networks
which parse a sentence into subsections. \cite{sutskever14seq2seq}
proposed RNNs that learn structures of long text sequences.%
{} Source codes of a program can also be seen as a text corpus with
its own grammar structure, thus being applicable for such neural network
models. \cite{karpathy15rnn} showed that a character-level LSTM trained
with Linux kernel codes is capable of detecting features such as brackets
or sentence length as well as generating simulated codes that greatly
resemble actual ones in syntactic structure. Motivated by such results
in the pattern discovery of source codes, several approaches have
been taken to solve practical issues in source code analysis. \cite{pu16skp}
and \cite{bhatia16auto} gathered data from programming assignments
submitted for a MOOC class to train a correction model which corrects
syntax errors in assignments. \cite{huo16learning} and \cite{lam16buggy}
applied attention-based CNN models to detect buggy source codes. \cite{allamanis14learning}
learned coding styles by searching for patterns with neural networks.
While these approaches proved that neural networks are capable of
detecting patterns within codes, they are limited to detecting only
syntax errors or bugs and not the transition of values stored inside
variables or functions of a source code program.

Neural networks with external memories have shown better performances
in inference or logical tasks compared to contemporary models. Following
the introduction of neural Turing machines~\cite{graves14ntm} and
memory networks~\cite{weston15mem,sukhbaatar15mem}, many variants
of these models were applied to various tasks other than QA tasks
such as sentiment analysis, part-of-speech tagging \cite{kumar16dmn},
and information extraction from documents \cite{miller16kv}. Yet,
so far there has been no work that applies a memory network-based
model to tasks with the complexity of semantic analysis in source
codes.

\section{Model Description}

\label{sec:model} 

In this section, we first provide the rationale for solving buffer
overruns as a QA task. We also introduce a source code-based training
dataset that we designed. Lastly, we describe the structure of our
model which is based on the memory network~\cite{sukhbaatar15mem}
and how it predicts buffer overruns from a source code.

\subsection{Benchmark Source Code Generation\vspace{-.1in}}

\begin{figure}[h]
\centering\subfloat[bAbI task example]{\centering\includegraphics[width=0.5\columnwidth,height=0.3\columnwidth]{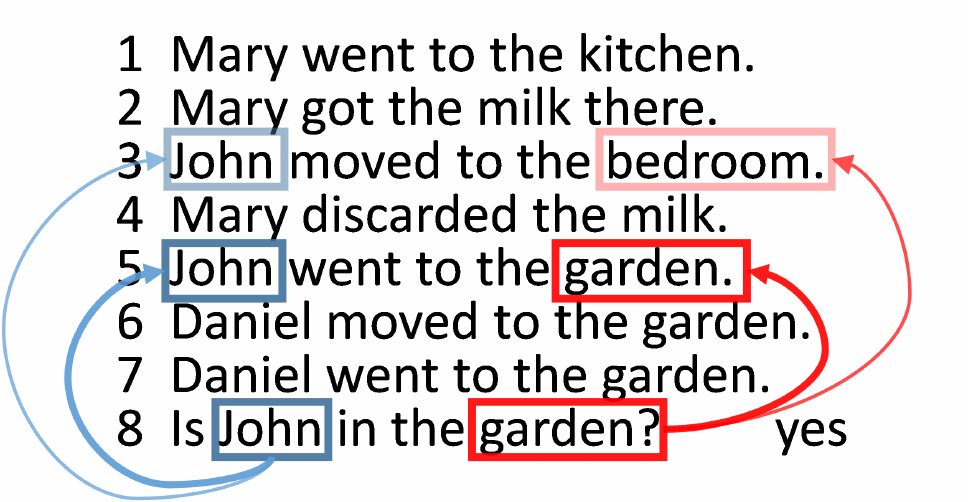}

\label{fig:babi_1} \vspace{0in}
}\subfloat[buffer overrun code sample]{\centering\includegraphics[width=0.5\columnwidth,height=0.3\columnwidth]{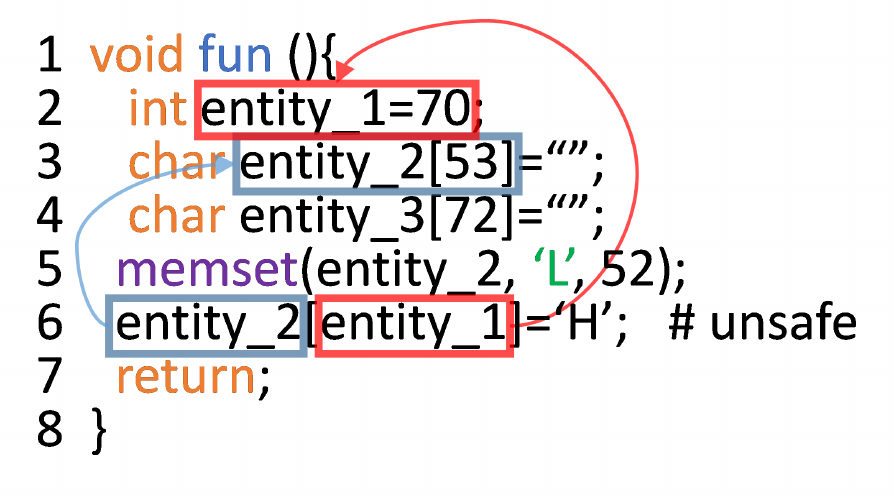}

\label{fig:source_code_1} \vspace{0in}
}\vspace{-.05in}\caption{Comparison of a bAbI and a buffer overrun tasks}

\label{fig:code_vs_babi}\vspace{-0in}
\end{figure}

\begin{figure*}[th]
\centering\subfloat[Level 1: direct buffer access]{\includegraphics[width=0.33\columnwidth]{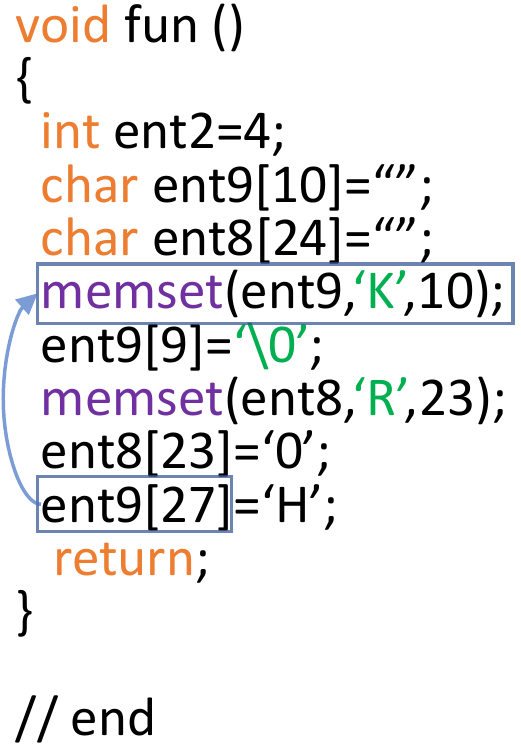}

\vspace{0in}
}\hspace{0.7in}\subfloat[Level 2: \textit{strcpy} access]{\includegraphics[width=0.33\columnwidth]{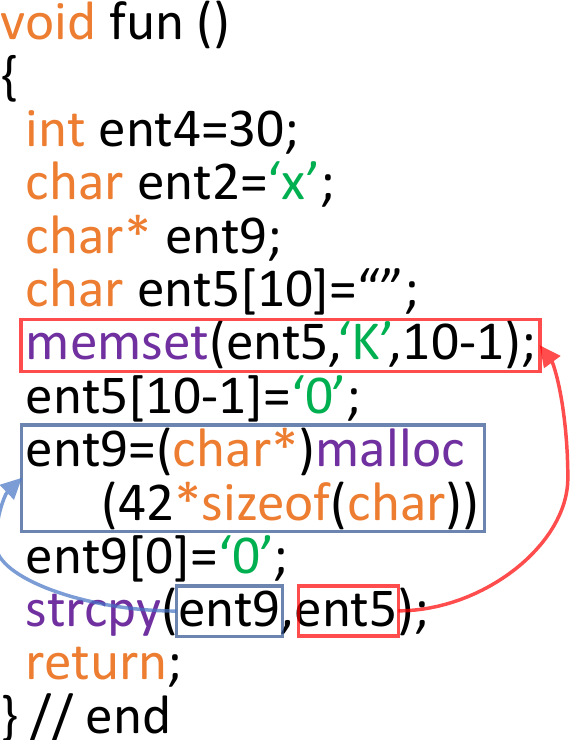}

\vspace{0in}
}\hspace{0.7in}\subfloat[Level 3: int allocation]{\includegraphics[width=0.33\columnwidth]{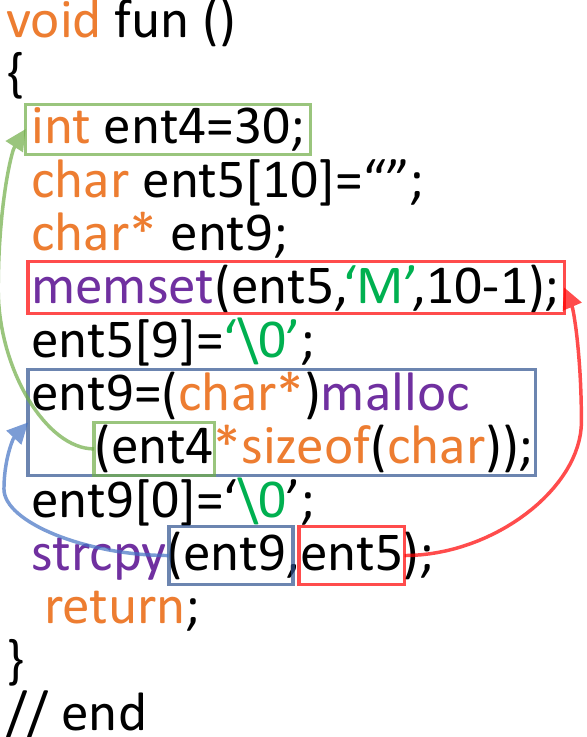}

\vspace{0in}
}\hspace{0.7in}\subfloat[Level 4: \textit{memcpy} access, int reallocation]{\includegraphics[width=0.33\columnwidth]{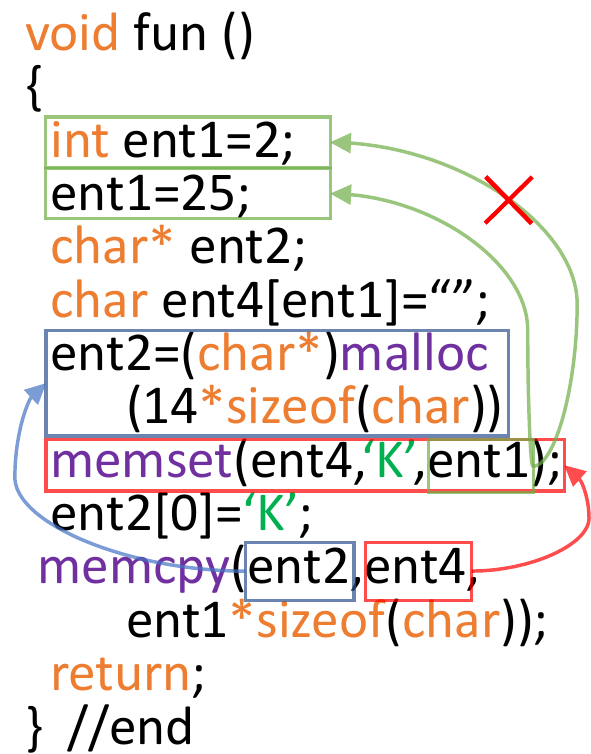}

\vspace{0in}
}

\vspace{-.1in}\caption{Different levels of buffer overrun tasks}
\vspace{-.1in}

\label{fig:levels}
\end{figure*}

\vspace{-.1in}\textbf{Comparison of bAbI tasks and buffer overruns}.
We return to our statement that analyzing buffer overruns is similar
to solving bAbI tasks. Consider Fig.~\ref{fig:code_vs_babi}. The
bAbI task shown in Fig.~\ref{fig:code_vs_babi}(a) is given a story
(lines 1-7) and a query (line 8). A solver model understands this
task by looking at `John' and `where' from the query and then attends
the story to find lines related to `John.' Lines 3 and 5 are chosen
as candidates. The model understands from the sequential structure
that line 5 contains more recent, thus relevant information. In the
end, the model returns the answer `garden' by combining the query
and the information from line 5.

Meanwhile, the task of Fig.~\ref{fig:code_vs_babi}(b) is to discriminate
whether the buffer access made at line 6 is valid. Our analyzer first
understands that its objective is to compare the size of the character
array \textit{entity\_2} and the integer variable \textit{entity\_1}.
Next, it searches for the length of \textit{entity\_2} at line 3,
where 53 is allocated to the variable. It also gains knowledge from
line 2 that \textit{entity\_1} is equivalent to 70. The remaining
task is to compare the integer variables 53 and 70 and return an alarm
(\textit{unsafe}) if the index exceeds the length of the character
array. One can think of lines 1-5 as a story and line 6 as a query,
perfectly transforming this problem into a bAbI task.

\textbf{Limitations of test suites}. Although test suites such as
Juliet Test Suite for C programming language~\cite{boland12juliet}
are designed for benchmarking buffer overrun and other program analysis
tasks, the data is not diverse enough. Code samples differ by only
a small fraction such as a different variable nested in a conditional
statement or loop, while a large portion of code appears repeatedly
over several samples. A data-driven model will inevitably learn from
only the small variations and ignore a large portion of the code where
much of the valuable information is stored.

\textbf{Program structure}. We tackle this problem of data inadequacy
by generating our own training source code dataset.\footnote{The generated dataset and generator codes are available at \url{https://github.com/mjc92/buffer_overrun_memory_networks}}
Our dataset adopts buffer access functions and initialization methods
from Juliet to maintain at least an equal level of task complexity,
while also preserving an underlying structure that makes it applicable
for deep learning approaches. Each sample is a void function of 10
to 30 lines of C code and consists of three stages: initialization,
allocation, and query. During the initialization stage, variables
are initialized as either characters, character arrays, or integers.
At the allocation stage, these variables are assigned values using
randomly generated integers between 1 to 100. Buffer sizes are allocated
to character arrays with \textit{malloc} and \textit{memset} functions.
At the query stage, a buffer access is attempted on one of the allocated
character arrays via a direct access on an array index (Fig.~\ref{fig:levels}(a)).%
{} We formulate this task into a binary classification problem where
an `unsafe' sign is returned if a character array is accessed with
a string or index that exceeds its size.

\begin{figure*}[t]
\centering\includegraphics[width=0.8\paperwidth]{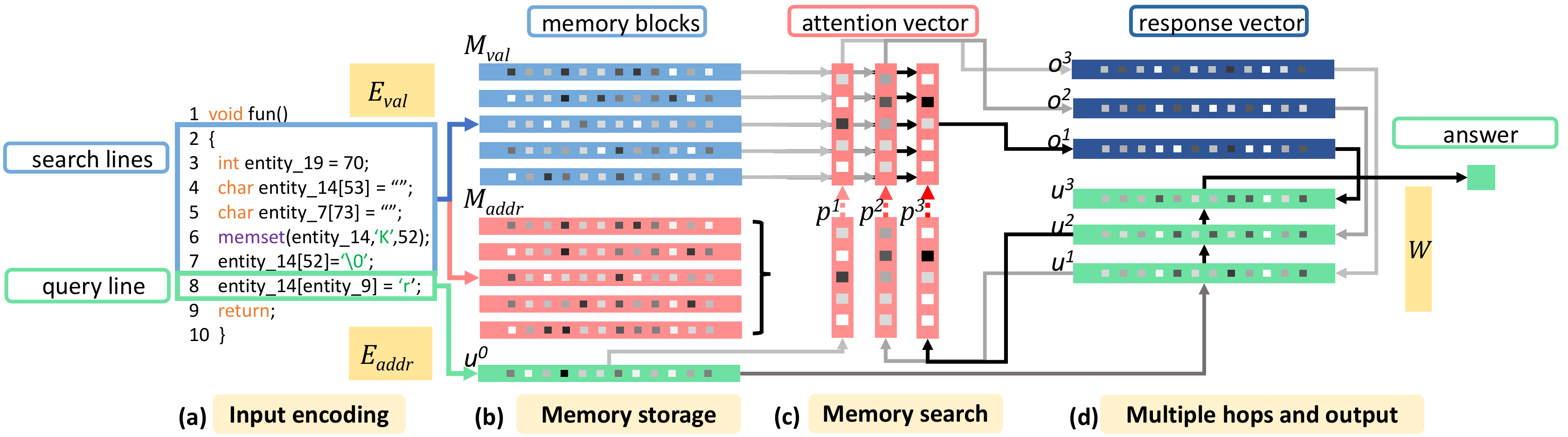}

\vspace{-.05in}\caption{Our proposed memory network-based model for buffer overrun tasks}

\label{fig:model}\vspace{-.05in}
\end{figure*}

\textbf{Naming conventions}. We assume that a limited number of individual
variables appear in each program sample. Each variable is given the
name\textit{ entity\_n} where $n\in\left\{ i|0\leqq i\leqq N_{upper},\,i\in Z\right\} $
and \textit{$N_{upper}$} is an integer set by default to 10. Each
$n$ is assigned randomly to variables and invariant of their introduced
order or data type. One can imagine a situation where an agent~(variable)
is given a fake ID~(entity name) for a particular task~(sample).
The agent learns to complete the task with that fake ID, then discards
it upon task completion, and selects a new one for the subsequent
task. In this manner, we can prevent \textit{entities} from learning
task-specific knowledge and instead train them as representations
of universal variables which can replace any kind of variable that
appears in a program. We can easily apply our model to real-life source
codes using this naming convention by simply changing the names of
newly introduced variables and methods to different \textit{entities}.

\textbf{Adding complexity}. Our model has to adapt to more realistic
source codes with complex structures. Possible settings that complicate
our task include
\begin{itemize}
\item selecting only the appropriate variables out of several dummy variables,\vspace{-.05in}
\item introducing different buffer access methods requiring the comparison
of two character arrays such as \textit{strcpy }or\textit{ memcpy}
functions,\vspace{-.05in}
\item allocating the sizes of character arrays not with integers but indirectly
with previously assigned integer variables,\vspace{-.05in}
\item reallocating integer variables prior to or after their use in allocating
a character array.
\end{itemize}
We first assign a number of dummy variables to each sample program.
Each dummy variable is initialized and allocated in the same manner
as the ones actually used in the buffer access. We include the use
of \textit{strcpy }(Fig.~\ref{fig:levels}(b)) / \textit{memcpy}
(Fig.~\ref{fig:levels}(d)) functions for buffer accesses. We also
add cases where character arrays are allocated not directly by integers,
but indirectly with additionally introduced integer variables (Fig.~\ref{fig:levels}(c)).
Given this setting, the model has to learn to store the integer value
allocated to the integer variable first, then use that value to obtain
the length of the character array. Finally, we add further cases where
the additional integer variable \textit{itself} is reallocated (Fig.~\ref{fig:levels}(d)),
either before or after it is used to define the character array length.
Now the model has to learn to choose whether the previously assigned
or reallocated value was used for allocating a character array.

Our generated source codes are equivalent to an expanded version of
Juliet test suite in terms of data flow, that is, the flow of data
values defined and used within a source code. Compared to codes in
Juliet which only define the source and destination variables that
will be used for buffer access, ours include dummy variables which
are defined and used similarly. The inclusion of various settings
such as memory allocation using assigned integer variables instead
of raw integers and reassignment of variables also increase data flow.
In overall, our source codes provide a tougher environment for models
to solve buffer overrun tasks than the existing Juliet dataset as
there are more variations to consider.  

\subsection{Model Structure}

The overall structure of our model is displayed in Fig.~\ref{fig:model}.

\textbf{Input encoding~(Fig.}~\ref{fig:model}\textbf{(a))}. The
memory network takes in as input a program code $X$ consisting of
$n$ search lines $X_{1},X_{2},\cdots,X_{n}$ and a single buffer
access line or query $X_{q}$. A single program line $X_{m}$ is a
list of words $w_{m}^{1},w_{m}^{2},\cdots,w_{m}^{l}$. With $V$ as
the vocabulary size, we define $x_{m}^{l}$ as the $V$-dimensional
one-hot vector representation of a word $w_{m}^{l}$. We set an upper
limit $N$ for the max number of lines a memory can store, and we
pad zeros for the remaining lines if a program is shorter than $N$
lines.

Note that every word in the source code is treated as a word token.
This includes not only variable names ($entity$), type definitions
($int$) and special characters, $\left(`[',\,`*'\right)$, but also
integers as well. This setting matches our concept of an end-to-end
model that does not require explicit parsing. While it is possible
to apply parsers to extract numbers and represent them differently
from other word tokens, this would contradict our goal of applying
a purely data-driven approach. Treating integers as individual word
tokens means that our model will not be given any prior information
regarding the size differences between numbers, and thus has to learn
such numerical concepts by itself. We further discuss this in Section
\ref{sec:experiments}.

Next, we compute vector representations for each sentence using its
words. Each word is represented in a $d$-dimensional vector using
an embedding matrix $E_{val}\in R^{d\times V}$. We also multiply
a column vector $l_{j}$ to the $j$-th word vector for each word
to allocate different weights according to word positions. This concept
known as position encoding~\cite{sukhbaatar15mem} enables our model
to discriminate the different roles of variables when two or more
identical words appear in a single sentence. Without such settings,
our model may fail to discriminate between source and destination
variables such as in a \textit{strcpy }function. The memory representation
$m_{i}$ of line $i$ consisting of $J$ words and the $k$-th element
$l_{j}^{k}$ of the position encoding vector $l_{j}\in\mathbb{R}^{d}$
for word $j$ in the line $i$ are obtained as \vspace{-.0in}
\begin{equation}
m_{i}=\Sigma_{j=1}^{t}l_{j}\cdot Ax_{i}^{j},
\end{equation}
\vspace{-.1in}
\begin{equation}
l_{j}^{k}=\left(1-j/J\right)-\left(k/d\right)\left(1-2j/J\right),
\end{equation}
 where `$\cdot$' is element-wise multiplication.

\textbf{Memory storage~(Fig.}~\ref{fig:model}\textbf{(b))}. Next,
we allocate our encoded sentences $m_{i}$ into matrices called memory
blocks. Fig.~\ref{fig:model}(b) shows two memory blocks, the memory
value block~$\left(M_{val}\in R^{N\times d}\right)$ and the memory
address block~$\left(M_{addr}\in R^{N\times d}\right)$. Each sentence
is allocated into one row of memory block, namely a memory slot. $M_{val}$
stores semantical information about the contents of a code line while
$M_{addr}$ stores information for locating how much to address each
line. For this reason, sentences are encoded using two different word
embedding matrices, $E_{val}$ and $E_{addr}$ for $M_{val}$ and
$M_{addr}$, respectively.

\begin{table*}[t]
\caption{Comparison on generated source codes. Inside brackets are the standard
deviations}
\centering{\scriptsize{}}%
\begin{tabular}{|c|ccc|ccc|ccc|ccc|}
\hline 
\multirow{2}{*}{\begin{turn}{90}
\end{turn}} & \multicolumn{3}{c|}{{\footnotesize{}level 1}} & \multicolumn{3}{c|}{{\footnotesize{}level 2}} & \multicolumn{3}{c|}{{\footnotesize{}level 3}} & \multicolumn{3}{c|}{{\footnotesize{}level 4}}\tabularnewline
\cline{2-13} 
 & {\footnotesize{}acc} & {\footnotesize{}F1} & {\footnotesize{}auc} & {\footnotesize{}acc} & {\footnotesize{}F1} & {\footnotesize{}auc} & {\footnotesize{}acc} & {\footnotesize{}F1} & {\footnotesize{}auc} & {\footnotesize{}acc} & {\footnotesize{}F1} & {\footnotesize{}auc}\tabularnewline
\hline 
\hline 
\multirow{2}{*}{{\footnotesize{}CNN}} & {\scriptsize{}0.67} & {\scriptsize{}0.69} & {\scriptsize{}0.75} & {\scriptsize{}0.73} & {\scriptsize{}0.71} & {\scriptsize{}0.81} & {\scriptsize{}0.61} & {\scriptsize{}0.61} & {\scriptsize{}0.66} & {\scriptsize{}0.62} & {\scriptsize{}0.62} & {\scriptsize{}0.67}\tabularnewline
 & {\scriptsize{}(0.01)} & {\scriptsize{}(0.02)} & {\scriptsize{}(0.01)} & {\scriptsize{}(0.02)} & {\scriptsize{}(0.02)} & {\scriptsize{}(0.01)} & {\scriptsize{}(0.01)} & {\scriptsize{}(0.03)} & {\scriptsize{}(0.01)} & {\scriptsize{}(0.01)} & {\scriptsize{}(0.03)} & {\scriptsize{}(0.01)}\tabularnewline
\hline 
\multirow{2}{*}{{\footnotesize{}LSTM}} & {\scriptsize{}0.8} & {\scriptsize{}0.84} & {\scriptsize{}0.92} & {\scriptsize{}0.82} & {\scriptsize{}0.80} & {\scriptsize{}0.90} & {\scriptsize{}0.69} & {\scriptsize{}0.66} & {\scriptsize{}0.76} & {\scriptsize{}0.67} & {\scriptsize{}0.64} & {\scriptsize{}0.75}\tabularnewline
 & {\scriptsize{}(0.01)} & {\scriptsize{}(0.01)} & {\scriptsize{}(0.00)} & {\scriptsize{}(0.01)} & {\scriptsize{}(0.02)} & {\scriptsize{}(0.01)} & {\scriptsize{}(0.00)} & {\scriptsize{}(0.01)} & {\scriptsize{}(0.01)} & {\scriptsize{}(0.01)} & {\scriptsize{}(0.02)} & {\scriptsize{}(0.01)}\tabularnewline
\hline 
{\footnotesize{}~Memory~} & {\scriptsize{}0.84} & {\scriptsize{}0.84} & {\scriptsize{}0.92} & {\scriptsize{}0.86} & {\scriptsize{}0.85} & {\scriptsize{}0.93} & {\scriptsize{}0.83} & {\scriptsize{}0.83} & {\scriptsize{}0.90} & {\scriptsize{}0.82} & {\scriptsize{}0.82} & {\scriptsize{}0.90}\tabularnewline
{\footnotesize{}~network~} & {\scriptsize{}(0.01)} & {\scriptsize{}(0.01)} & {\scriptsize{}(0.01)} & {\scriptsize{}(0.01)} & {\scriptsize{}(0.02)} & {\scriptsize{}(0.01)} & {\scriptsize{}(0.02)} & {\scriptsize{}(0.02)} & {\scriptsize{}(0.02)} & {\scriptsize{}(0.02)} & {\scriptsize{}(0.02)} & {\scriptsize{}(0.02)}\tabularnewline
\hline 
\end{tabular}{\scriptsize{}\vspace{-.05in}}{\scriptsize \par}

\label{tab:scores}
\end{table*}

\textbf{Memory search~(Fig.}~\ref{fig:model}\textbf{(c))}. The
query is encoded into a representation using $E_{addr}$. We denote
the initial query embedding as $u^{0}$. By computing the inner products
between the query embedding and each slot of the memory address block,
then applying a softmax function to the resulting vector, we obtain
the attention vector $p$ which indicates how related each line is
to the query. The $i$-th element of $p$ is obtained as\vspace{-.05in}
\begin{equation}
p_{i}=\text{softmax}\left(\left(u^{0}\right)^{T}M_{addr}\right),\label{eq:attention}
\end{equation}

\vspace{-.05in}with the softmax function as \vspace{-.05in}
\begin{equation}
\text{softmax}\left(z_{i}\right)=e^{z_{i}}/\Sigma_{j}e^{z_{i}}.
\end{equation}

\vspace{-.05in}The response vector $o$ is computed as in \vspace{-.05in}
\begin{equation}
o=\Sigma_{i}p_{i}\left(M_{val}\right)_{i}.\label{eq:o}
\end{equation}

\vspace{-.05in}This vector contains information collected over all
lines of the memory value block according to their attention weights
obtained from the memory address block. This is equivalent to searching
the memory for different parts of information with respect to a given
query.

\textbf{Multiple hops and output~(Fig.}~\ref{fig:model}\textbf{(d))}.\textbf{
}The response vector $o$ can be either directly applied to a weight
matrix $W$ to produce an output, or added to strengthen the query
$u$. In the latter case, the query is updated as in Eq.~(\ref{eq:update})
by simply adding the response vector to the previous query embedding.
\begin{equation}
u^{k+1}=u^{k}+o^{k}\label{eq:update}
\end{equation}

\vspace{-.05in}We repeat from Eq.~(\ref{eq:attention}) to obtain
a new response vector. Our model iterates through multiple hops where
at each hop the desired information to be obtained from the memory
slightly changes. This accounts for situations where a model has to
first look for lines where an array is allocated, and then gather
information from lines stating the size of the variables used for
allocating the array size. The final output is a floating value ranging
from 0 (unsafe) to 1 (safe), which we round to the nearest integer
to obtain a binary prediction result.

\section{Experiments}

\label{sec:experiments} 

In this section, we present both quantitative and qualitative results
of our experiments on model performance and learned characteristics.\vspace{-.05in}

\subsection{Quantitative Evaluation}

\vspace{-.05in}\textbf{Experiment settings}. Our main dataset consists
of C-style source codes discussed in Section~\ref{sec:model}.2.
We used a single training set consisting of 10,000 sample programs.
We generated four test sets with 1,000 samples each and assigned them
levels one to four, with a higher level indicating a more complex
condition (see Table~\ref{tab:levels}). Samples ranged from 8 to
33 lines of code, with an average of 16.01. A total of 196 unique
words appeared in the training set. A maximum of four dummy variables
were added to each sample. We used random integers between 0 and 100
for buffer allocation and access. We conducted every experiment on
a Intel~(R) Xeon~(R) CPU E5-2687W v3 @ 3.10GHz machine equipped
with two GeForce GTX TITAN X GPUs. All models were implemented with
Tensorflow 0.12.1 using Python 2.7.1 on an Ubuntu 14.04 environment.

\textbf{Model Comparison}. We set our memory network to three hops
with a memory of 30 lines and the embedding size of $d=32$. As there
has been no previous work on using deep learning models for such tasks,
we used existing deep learning models often used for text classification
tasks as baselines. That is, we included a CNN for text classification~\cite{kim14cnn}
and a two-layer LSTM binary classifier. All models were trained with
Adam~\cite{kingma14adam} at a learning rate of 1e-2. We used the
classification accuracy, F1 score, and the area under the ROC curve
(AUC) as performance metrics. We averaged the scores of the ten best
cases with the smallest training error.

\begin{table}[h]
\caption{Different levels of test sets}
\vspace{-.05in}\centering%
\begin{tabular}{|c|cccc|}
\hline 
 & {\scriptsize{}Level 1} & {\scriptsize{}Level 2} & {\scriptsize{}Level 3} & {\scriptsize{}Level 4}\tabularnewline
\hline 
{\scriptsize{}Direct buffer access} & {\footnotesize{}$\surd$} & {\footnotesize{}$\surd$} & {\footnotesize{}$\surd$} & {\footnotesize{}$\surd$}\tabularnewline
{\scriptsize{}Access by}\textit{\scriptsize{} strcpy}{\scriptsize{}
/}\textit{\scriptsize{} memcpy} &  & {\footnotesize{}$\surd$} & {\footnotesize{}$\surd$} & {\footnotesize{}$\surd$}\tabularnewline
{\scriptsize{}Allocation by int variable} &  &  & {\footnotesize{}$\surd$} & {\footnotesize{}$\surd$}\tabularnewline
{\scriptsize{}Reallocation of int variable} &  &  &  & {\footnotesize{}$\surd$}\tabularnewline
\hline 
\end{tabular}\vspace{-.1in}

\label{tab:levels}
\end{table}

Performance results shown in Table~\ref{tab:levels} demonstrate
that all models decrease in performance as task levels increase, due
to our level assignment. Of the deep learning models, only memory
networks performed consistently at a high level on all four level
settings with accuracy rates higher than 80\%. This is expected since
their hops allow them to solve even complex situations such as variable
reallocation and different buffer access types. Meanwhile, CNNs failed
to complete even the simplest tasks since they cannot capture the
sequential information in input sentences and instead apply convolutional
filters to words of all regions on an equal basis. Any positional
information is discarded. 

Memory networks require substantially shorter computing time compared
to other models, requiring an average of 0.63s per epoch, while LSTMs
and CNNs each require 7.13s and 13.21s. As for the latter models,
the number of computations is proportional to the number of words
that appear in a code. However, memory networks sum up all words of
a sentence to form a single representation, and thus computation time
relies on the number of lines instead of individual words. This significantly
reduces computation time.

\begin{figure*}[t]
\centering\subfloat[Cosine similarities]{\includegraphics[width=0.5\columnwidth,height=0.3\columnwidth]{data/figures/figure_embedding_visualization/number_embedded_cosine}}\subfloat[L2-norm distances]{\includegraphics[width=0.5\columnwidth,height=0.3\columnwidth]{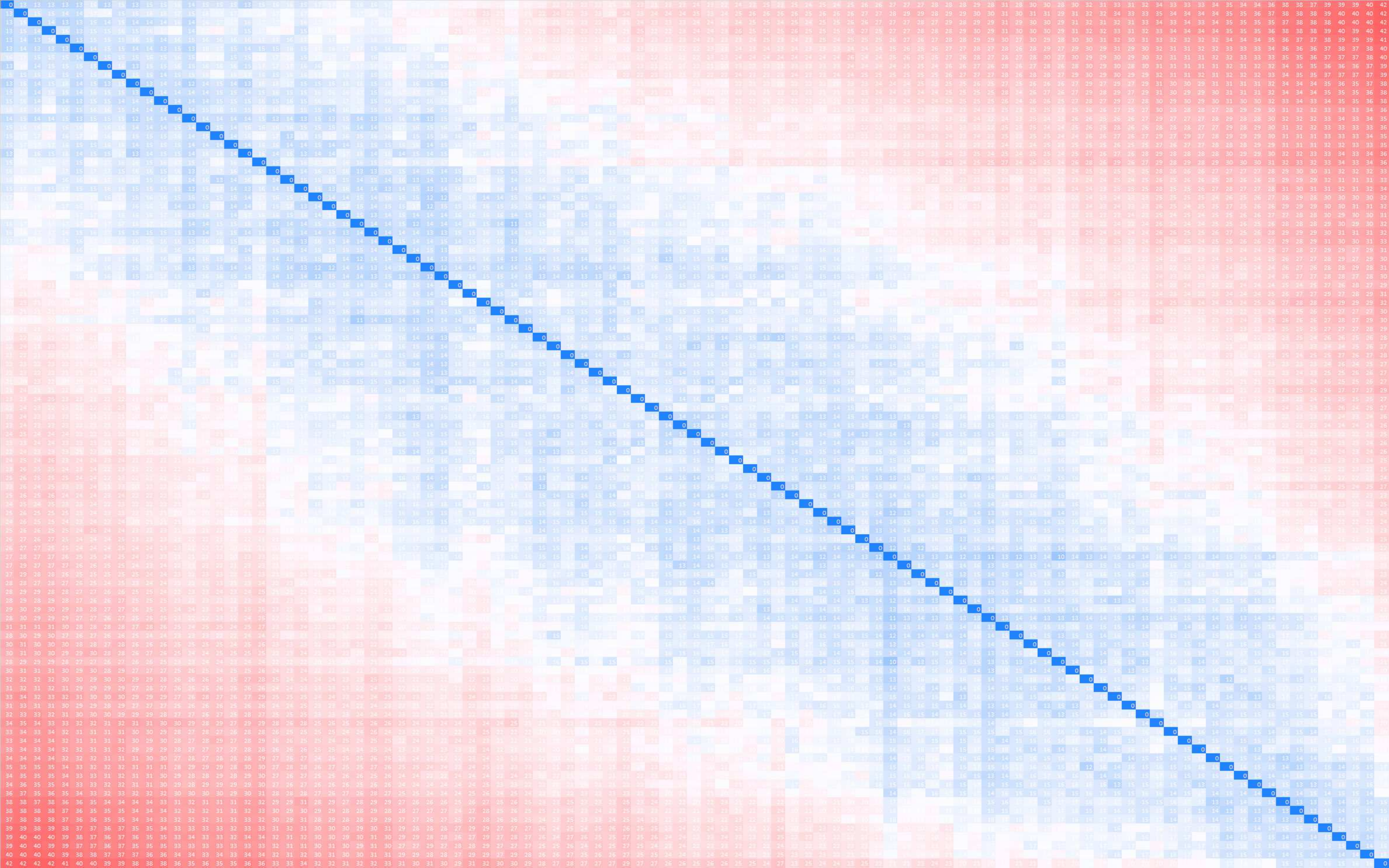}}\subfloat[Representations of learned word embeddings]{\centering\includegraphics[width=0.5\columnwidth,height=0.3\columnwidth]{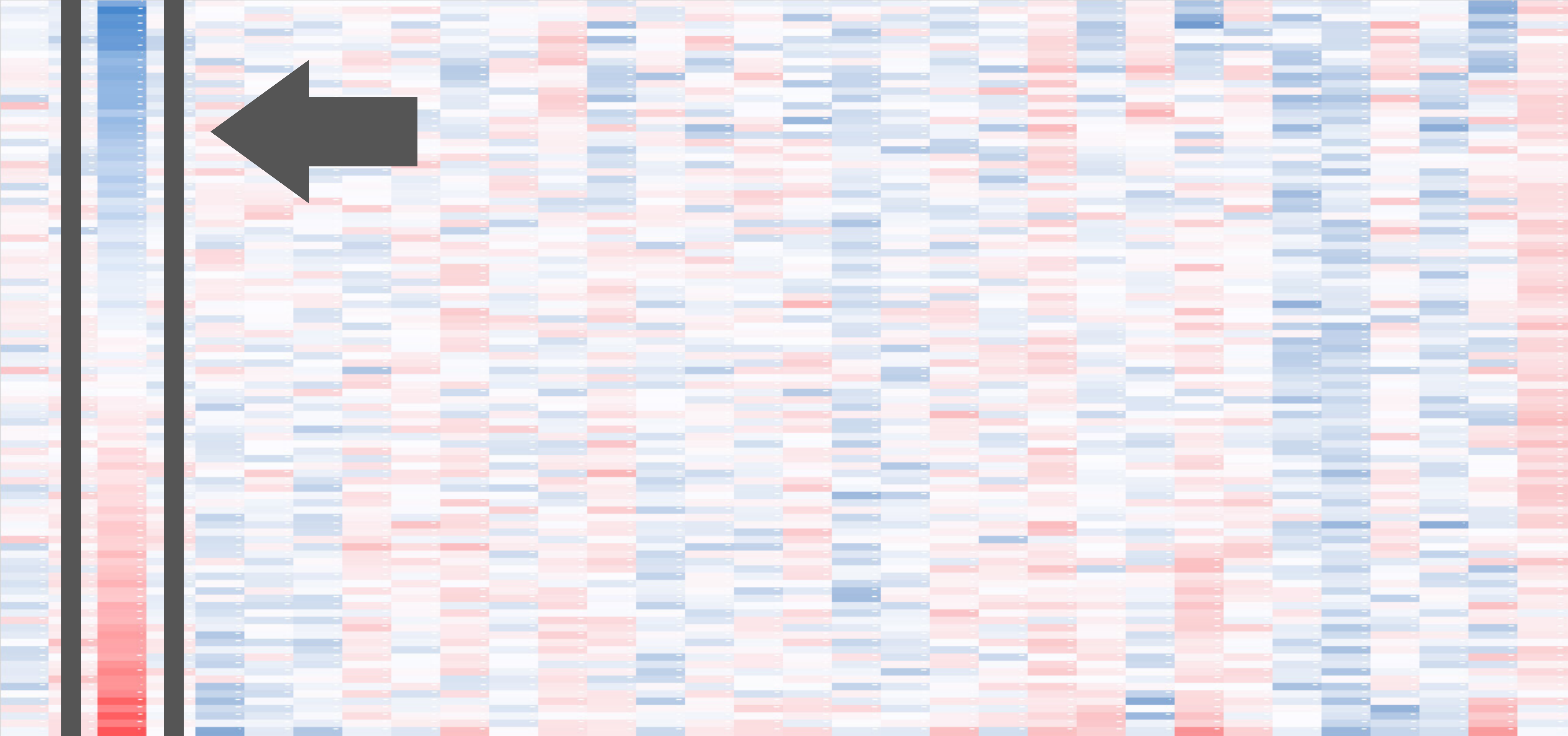}}\subfloat[Visualization with t-SNE]{\centering\includegraphics[width=0.5\columnwidth,height=0.3\columnwidth]{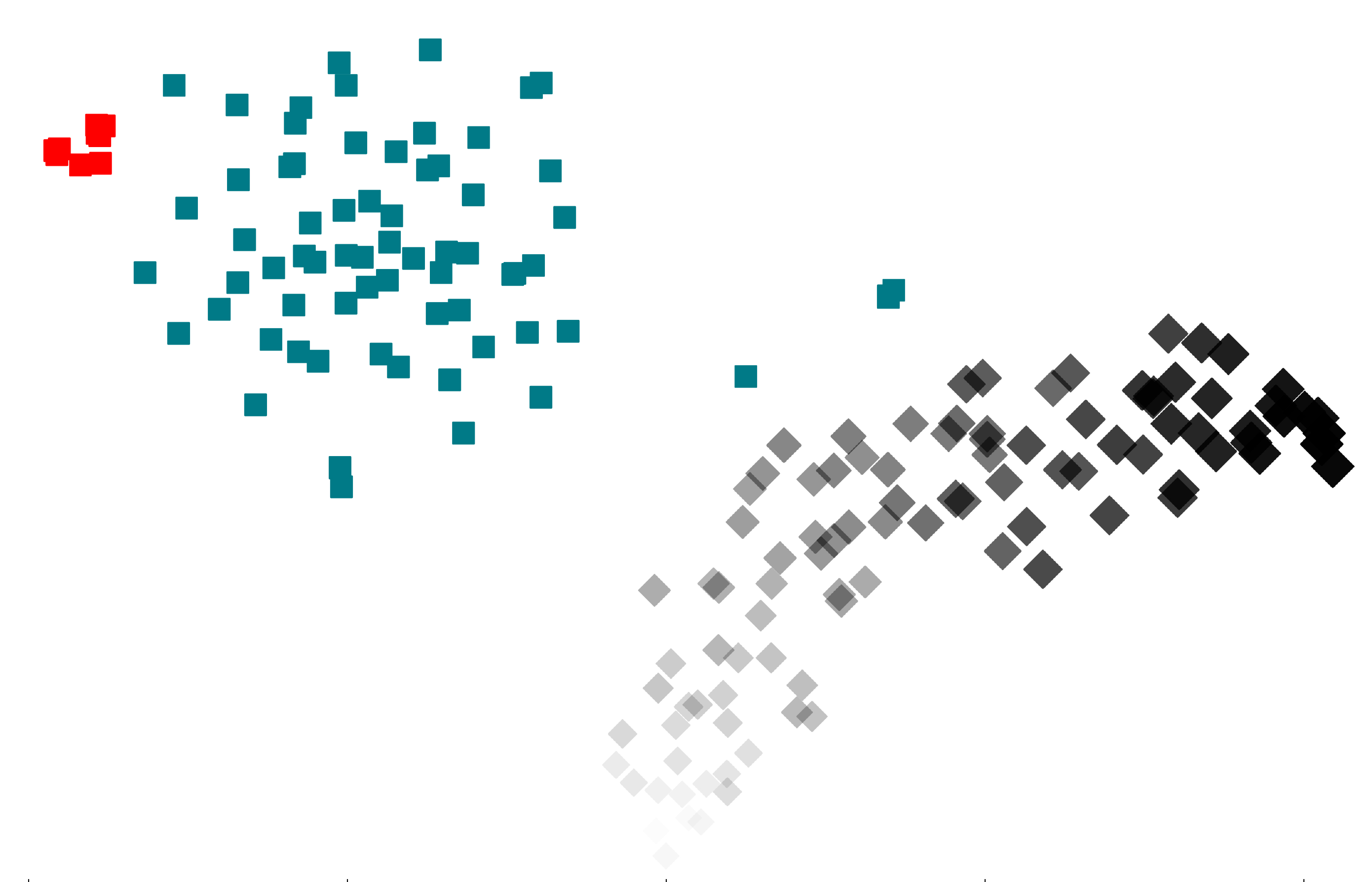}}\caption{Visualizations of word embedding vectors of numbers 1-100. Red and
blue indicate high and low values, respectively. }

\label{fig:vis-1}
\end{figure*}

Interestingly, LSTM models also performed well when set to easier
tasks. Results show that LSTMs performed comparably to memory networks,
even equaling them on Level 1 tasks. However, its performance sharply
dropped when used on higher level tasks. This partial success of LSTMs
relates to the simple structure of Level 1 tasks. The size of the
character array always appears before the index to access, so the
model can cheat by comparing the only two numbers that appear within
the entire code. This cheating becomes obsolete as higher-level tasks
require knowledge only obtainable by attending previous lines in a
stepwise manner.%

\subsection{Qualitative Analysis}

\begin{figure}[h]
\centering\includegraphics[width=1\columnwidth]{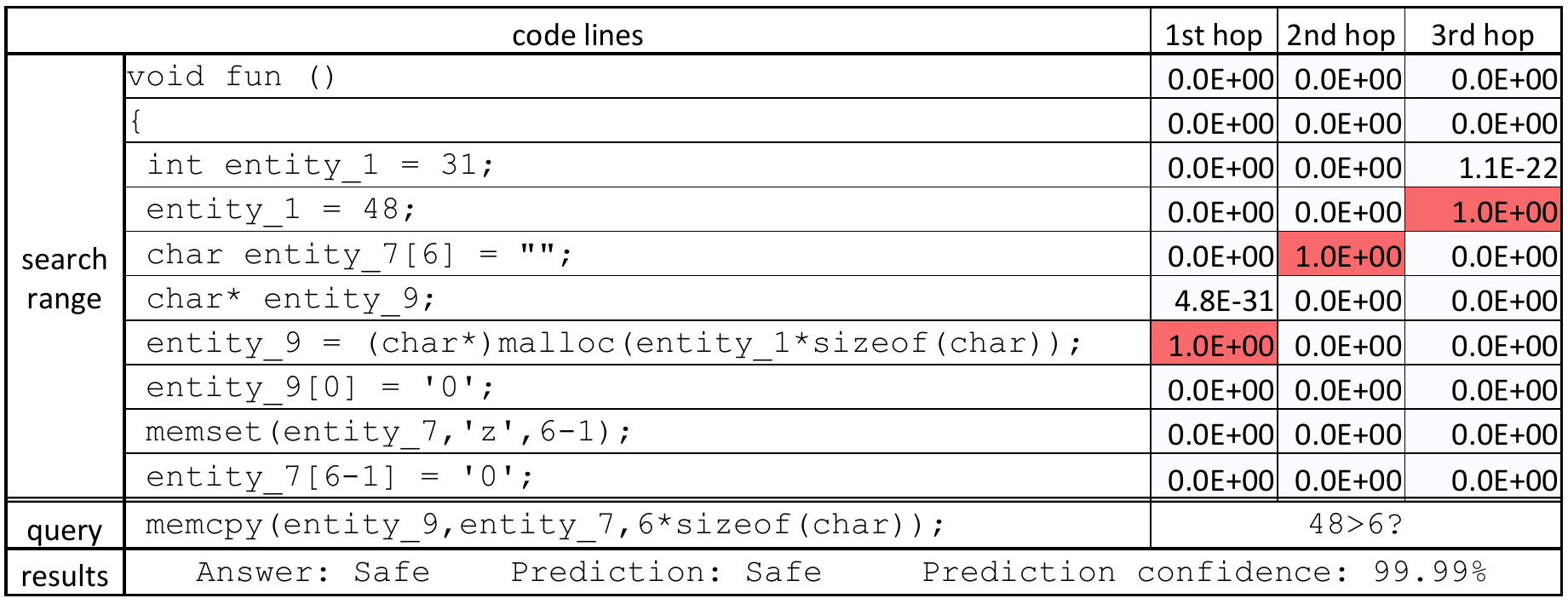}

\caption{Prediction result with attention per hop}
\vspace{-0.2in}

\label{fig:pred-results-1}
\end{figure}

We further examine the performance of our memory network model and
the steps it takes to obtain a correct answer. We also present visualization
results on how our model learns the concepts of numbers and numerical
comparison without being explicitly supervised about such tasks. 

\textbf{Tracking hop-wise results.} In order to prove that our model
solves the tasks in our desired manner, that is, by attending and
collecting relevant information from different parts of the memory
at different hops, we analyze individual prediction cases by inspecting
which parts of information our model has obtained from taking each
hop.

Fig.~\ref{fig:pred-results-1} displays an example of buffer overrun
analysis using our model. We can observe that when given a \textit{strcpy}
buffer access as a query, the model's initial attention shifts to
the sentence where the destination buffer (\textit{entity\_3}) is
allocated. The model decides here to next look for \textit{entity\_9},
which contains the size used for allocating to \textit{entity\_3}.
During the next hop it attends the line where the source buffer (\textit{entity\_2})
is allocated and obtains data of 99, the size of \textit{entity\_2}.
At the last hop the memory network visits entity\_9 and obtains 69.
After the three hops, the destination size 69 is compared with source
size 99, and being a smaller number, returns `unsafe' as a result.
The prediction confidence in Fig.~\ref{fig:pred-results-1} indicates
how close the predicted value is to the ground answer.

\textbf{Numerical concepts automatically learned.} Recall from Section~\ref{sec:model}
that our model was not given any prior information regarding the notion
of quantitative values. Interestingly, our model learned to compare
between different numbers. Fig.~\ref{fig:vis-1} displays visualization
results using only the word embedding vectors corresponding to the
100 numbers.

Figs.~\ref{fig:vis-1}(a) and (b) display the cosine similarities
and the L2-norm distances of all numbers from 1 to 100, with 1 at
the topmost left-hand side. The colors observed at the first and third
quadrants from both figures show that numbers with large differences
are trained to minimize cosine similarities while maximizing L2-norm
distances, thus spacing themselves apart. In contrast, similar numbers
in the second and fourth quadrants have opposite characteristics,
meaning they are similarly placed.

The word embedding vectors of numbers across all $d$ dimensions as
seen in Fig.~\ref{fig:vis-1}(c) further demonstrate a clear sequential
order between numbers. The highlighted column forms a strong color
spectrum starting from a low value which gradually increases as the
corresponding number increases from 1 to 100. As all word embeddings
were initialized with random values at the beginning, this spectrum
indicates that our model learns by itself to assign such values for
comparison purposes.%
{} 

Last of all, Fig. \ref{fig:vis-1}(d) is a t-SNE representation of
all word embedding vectors. The black gradation indicates the word
embeddings of numbers, with denser colors indicating larger numbers.
We notice that they are embedded in a consistent direction in an increasing
order. While this again shows how our model learns numerical characteristics,
we also discover that dots in red, which correspond to \textit{entities}
from Section~\ref{sec:model}, stand out. As mentioned earlier, \textit{entities}
correspond to the variables that appear in source codes as integer
variables or character buffers. This implies that our model learns
to train word embeddings differently according to their purposes within
a code.

\vspace{-.05in}

\section{Conclusions and Future Work}

\label{sec:conclusion} 

In this work, we proposed a memory network-based model for predicting
buffer overruns in programming language analysis. Our work is the
first to apply a deep learning-based approach to a problem in the
field of program analysis that requires both syntactic and semantic
knowledge. Performance results show that memory networks are superior
to other models in solving buffer overrun tasks across all difficulty
levels. We also presented that our model successfully learns the notion
of numbers and their quantitative comparisons from merely textual
data in an end-to-end setting.

Our work has room to improve in many interesting aspects. We can expand
our model to cover different program analysis tasks such as pointer
analysis, interval analysis, and flow-sensitivity analysis, which
share similar semantic natures. We can apply advanced variants of
memory networks to handle various conditions in source codes such
as \textit{if} and \textit{for} statements. Our knowledge of models
learning numerical representations can further aid deep learning models
compatible with arithmetic and logical reasoning. All of these combined,
our work marks a stepping stone to a fully data-driven program analyzer.

{\small{}\bibliographystyle{named}
\bibliography{my_ijcai,sehun_ijcai}
}{\small \par}
\end{document}